\documentclass[12pt]{iopart}
\usepackage{iopams}  
\usepackage{graphicx}

\begin{document}

\title[Planar Channelling: Monte-Carlo Simulations]{Planar Channelling of 855 MeV Electrons in Silicon: Monte-Carlo Simulations}

\author{Andriy Kostyuk$^{1}$, Andrei Korol$^{1,2}$, Andrey Solov'yov$^{1,3}$, and Walter Greiner$^{1}$}

\address{$^{1}$ Frankfurt Institute for Advanced Studies, Johann Wolfgang 
Goethe-Universit\"at,
60438 Frankfurt am Main, Germany}

\address{$^{2}$ Department of Physics, St. Petersburg State Maritime 
Technical University,
198262 St. Petersburg, Russia}

\address{$^{3}$ A.F. Ioffe Physical-Technical Institute, 194021 St. Petersburg, Russia}

\begin{abstract}
A new Monte Carlo code for the simulation of the channelling of
ultrarelativistic charged projectiles in single crystals is presented.
A detailed description of the underlying physical model and the 
computation algorithm is given.
First results obtained with the code for the channelling of 855 MeV electrons 
in Silicon crystal are presented. The dechannelling lengths for (100), (110) and (111)
crystallographic planes are estimated.
In order to verify the code, the dependence of the intensity of the channelling 
radiation on the crystal dimension
along the beam direction is calculated. A good agreement of the obtained results  
with recent experimental data is observed.
\end{abstract}

\submitto{\jpb}
\maketitle

\section{Introduction}

In this article we consider planar channelling of 855 MeV electrons in Silicon crystal using
a new Monte-Carlo code.

Channelling takes place if charged particles enter a single crystal at small
angle with respect to crystallographic planes or axes \cite{Lindhard}. The 
particles get confined
by the interplanar or axial potential and move preferably along the 
corresponding crystallographic planes or axes following their shape.

Recent revival of the interest to this phenomenon is due to its growing 
practical application. In particular, the crystals with bent crystallographic 
planes are used to steer high-energy charged particle beams 
replacing huge dipole magnets.
Since its appearance \cite{Tsyganov1976} and first experimental verification \cite{Elishev1979}
this idea  has been attracting a lot of interest worldwide. Bent crystal
have been routinely used for beam extraction in the Institute for High Energy Physics, Russia
\cite{Afonin2005}. A series of experiments on the bent crystal deflection and collimation of proton and heavy ion beams were
performed at different
accelerators \cite{Arduini1997,Carrigan1999,Fliller2006,Strokov2007,Scandale2008,Scandale2010zzb} throughout the world.
The bent crystal method has been proposed to extract particles from the beam halo at
CERN's Large Hadron Collider \cite{Uggerhoj2005}.
The possibility of deflecting positrons \cite{Bellucci2006}, electrons \cite{Strokov2007,Strokov2006,Scandale2009zz}
and $\pi^{-}$ mesons \cite{Scandale2009zzb,Scandale2010} has been studied as well.

Another very promising application of the channelling phenomenon is a novel 
source of hard electromagnetic radiation.
A single crystal with periodically bent crystallographic planes can force 
channelling particles to move along  nearly sinusoidal trajectories
and radiate in the hard x- and gamma-ray frequency range.
The feasibility of such a device, known as the 'crystalline undulator`,
was demonstrated theoretically a decade ago \cite{first,KSG1999} (further
developments as well as historical references are reviewed in \cite{KSG2004_review}).
The advantage of the crystalline undulator is in extremely strong
electrostatic fields inside a crystal, which are able
to steer the particles much more effectively than even the most advanced
superconductive magnets. 
This fact allows one to make the period $\lambda_u$ of the crystalline undulator
in the hundred or even ten micron range, which is two to three orders of 
magnitude smaller than that of conventional undulator. Therefore
the wavelength of the produced radiation 
$\lambda \sim \lambda_\mathrm{u}/(2 \gamma^2)$ ($\gamma \sim 10^3$--$10^4$ being the 
Lorentz factor of the particle) can reach the (sub)picometer range, 
where conventional sources with 
comparable intensity are unavailable \cite{Topics}.

Initially, it was proposed to use positron beams in the crystalline undulator.
Positively charged particles are repelled by the crystal nuclei and, therefore,
they move between the crystal planes, where there are no atomic nuclei and 
the electron density is less then
average. This reduces the probability of random collisions with the crystal 
constituents. Hence, the transverse momentum of the particle increases
slowly and the particle travels a longer distance in the channelling regime. 

More recently, an electron based crystalline undulator was proposed 
\cite{Tabrizi}. On one hand, electrons are less preferable than positrons.
Due to their negative charge, the electrons are attracted by the 
lattice ions and, therefore, are forced to oscillate around the crystal 
plane in the process of channelling. The probability of collisions with
crystal constituents is enhanced. Thus, the dechannelling length is smaller
by about two orders of magnitude in comparison to that of positrons at the same 
conditions. 
On the other hand, the electron beams are easier available and are usually
of higher intensity and quality. Therefore, from the practical point of view,
electron based crystalline undulator has its own advantages and deserves 
a thorough investigation.

There is another reason why electron channelling needs a thorough analysis.
This is the disagreement between theory and experimental data.
For example, the Baier-Katkov-Strakhovenko  formula for the dechannelling length 
(equation (10.1) in \cite{Baier_book}), $L_\mathrm{d}$, predicts 
$L_\mathrm{d}=23\ \mu$m for $1.2$ GeV electrons  in  Si (110)
planar channel, while the value extracted from experimental data is 
$L_\mathrm{d}=28\ \mu$m \cite{AdejshviliEtAl}.
At $855$ MeV, the formula yields $L_\mathrm{d}=15.7\ \mu$m vs 
$L_\mathrm{d}=18.\ \mu$m obtained from a model dependent 
analysis of the experimental data \cite{Backe2008}. These disrepencies are rather small 
and can be attributed to different definitions of the dechanneling length 
used by experimental and theoretical groups.

For lower energies, the discrepancy is much more 
dramatic: $L_\mathrm{d}=6.7\ \mu$m calculated vs. 
$L_\mathrm{d}=31\ \mu$m measured \cite{KomakiEtAl1984}
and $L_\mathrm{d} \approx 1\ \mu$m  calculated vs. $L_\mathrm{d}=36\ \mu$m  measured
\cite{KephartEtAl1989}
for electron energies $350$  MeV and $54$ MeV, respectively.

Clearly, further theoretical and experimental investigations of the electron channelling 
are necessary (see also \cite{Carrigan:2009}). 
No accurate theoretical description of the electron deflection by bent crystals or 
the electron-based crystalline 
undulator is possible until an adequate and experimentally verified theoretical or numerical model 
of electron channelling is available.

To build such a model,
we developed a new Monte-Carlo code that allows us to simulate the particle
channelling and calculate the emitted radiation. In contrast to other channelling
codes \cite{Artru1990,Biryukov1995,Bogdanov},
our algorithm does not use the continuous potential approximation. This novel 
feature is especially beneficial in the case of negatively charged projectiles, which
channel in the vicinity of the atomic nuclei, where the continuous potential approximation 
becomes less accurate.

In this paper we present the first results obtained with our code. We  have studied
the channelling of 855 MeV electrons in a straight single crystal of Silicon along three
different crystallographic planes: (100), (110) and (111). The parameters
of the simulation correspond to the conditions of the channelling experiments
at Mainz Microtron (Germany) \cite{Backe2008}. To verify our results, we calculated the
dependence of the intensity of the channelling radiation on the crystal dimension
along the beam direction and compared to the experimental data.

\section{The Underlying Physical Model and its Validity Domain}
\label{Model}

Our model is intended for studying the interaction of ultrarelativistic projectiles
with single crystals. It is best suited for light projectiles: electrons 
and positrons, but it can be also used for ultrarelativistic heavy projectiles.

\begin{figure}[htp]
\begin{center}
\includegraphics*[width=12cm]{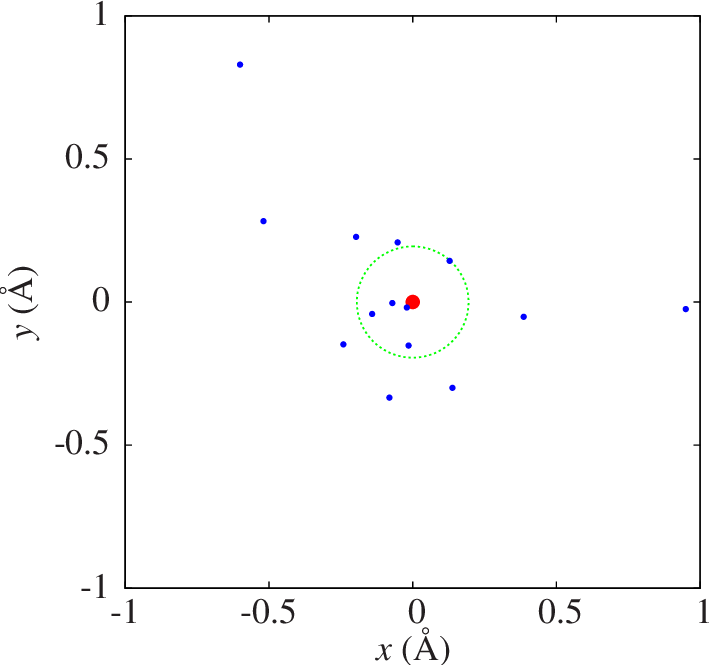}
\end{center}
\caption{An example of a `snapshot' of a Silicon atom as it is seen by an ultrarelativistic projectile
moving along the $z$-axis.
The larger and smaller circles represent the nucleus and the electrons, respectively. The dashed
line shows the Thomas-Fermi radius of the atom.}
\label{atom}
\end{figure}

Due to the high speed of the projectile, its interaction time with a crystal atom is 
very short. The atomic electrons have no time to move during the interaction.
As a result, the projectile `sees' a `snapshot' of the atom: the atomic electrons 
are seen as point-like charges at fixed positions around the nucleus (see Figure \ref{atom}).

The probability density to see the atomic electrons at positions 
$\vec{r}_{1}$, $\vec{r}_{2}$, $\dots$, $\vec{r}_{Z}$ ($Z$ is the atomic number) is given
by squared absolute value of the wave function of the atom:  
\begin{equation}
w(\vec{r}_{1}, \vec{r}_{2}, \dots, \vec{r}_{Z}) =
| \psi (\vec{r}_{1}, \vec{r}_{2}, \dots, \vec{r}_{Z}) |^{2} .
\end{equation}

Instead of using the exact wave function, we approximate it by a spherical symmetric probability 
distribution that on average reproduces the electrostatic potential of the atom in 
Moli\`ere's approximation \cite{Moliere}. Our approach ignores nonsphericity of electron 
orbitals as well as anticorrelations, due to Coulomb's repulsion and Pauli's principle,
between electron positions.

The interaction of the projectile with an atomic constituent is considered as a classical scattering in 
a Coulomb field of a static point-like charge. A projectile with electric 
charge $q_\mathrm{p}$ and initial speed $v$ along the $z$-axis attains after interaction with 
a static charge $q_{t}$ a transverse momentum
\begin{equation}
\Delta \vec{p}_{\perp} = - 2 q_\mathrm{p} q_\mathrm{t} \frac{\vec{r}_{\perp}}{v \vec{r}_{\perp}^{2}},
\label{pperp}
\end{equation}
$\vec{r}_{\perp}$ is the vector connecting the projections of the projectile and the static charge
onto the $(xy)$-plane.

The total transverse momentum attained by the projectile 
in the collision with the atom is a vector sum of (\ref{pperp}) over all atom constituents.
The absolute value of the projectile momentum 
remains unchanged. This means that the projectile energy losses for ionisation or excitation 
of the atom are neglected. Indeed,  
the ionization losses of high energy electrons or positrons 
in matter are very small (see, for example, Figure 27.10 in \cite{PDG}).

The above procedure is approximate. It is valid if the scattering angle 
$\vartheta = |\Delta \vec{p}_{\perp}|/p$ is small. In the opposite case, not only formula (\ref{pperp})
but also the representation of the atom as a collection of static charges become wrong.
However, accurate description of the large angle scattering is not
important for modelling of the channelling processes. First, such processes are rare.
Second, if the projectile is scattered by an angle that is much larger than the critical 
(Lindhard's) angle $\vartheta_\mathrm{L}$ \cite{Lindhard}, its probability to return to 
the channelling regime is 
negligible. Therefore the trajectory of a projectile after a large-angle scattering
is out of our interest 
and the precise value of the large scattering angle does not matter.

On the other hand, the scattering by $\vartheta \lesssim \vartheta_\mathrm{L}$
is important for a proper modelling of the channelling 
phenomenon. Lindhard's angle for ultrarelativistic projectiles is typically in the 
submiliradian range. The described procedure 
as well as formula (\ref{pperp}) are valid for such small scattering angles.

The `snapshot' model is applied not only to each atom but also to the crystal as a
whole. The thermal motion of the atoms is even slower than the motion of atomic electrons.
Therefore, the projectile sees the atomic nuclei `frozen' at random positions in 
the vicinity of nodes of the crystal lattice. The probability distribution of the position
of the nucleus relative to the node can be approximated by a three dimensional normal 
distribution with the variance equal to the squared amplitude of 
thermal vibrations of the crystal atoms.

Although the crystal constituents are considered as static point-like 
charges,
and the model looks like completely classical construction at the first sight, the quantum 
properties of the crystal are properly taken into account: the probability distributions of 
electrons and nuclei are found from the quantum theory.

In contrast, the quantum aspects
of the projectile motion are indeed completely ignored. Our code performs calculation of 
trajectories of the projectiles in the crystal.
Calculation of a particle trajectory implies that its coordinate and 
momentum can be measured simultaneously. 
Let us estimate the validity domain of this 
approximation.

The notion of trajectory makes sense in the case of a channelling particle if one is 
able to determine the particle coordinate $y$ and the transverse momentum $p_{y}$ 
with accuracies $\delta \! y$ and $\delta \! p_{y}$ that are much smaller than the 
channel width $d$ and typical value $p_{y}^\mathrm{ch}$ of the transverse momentum of 
a channelling particle. On the other hand, the product $\delta \! p_{y}  \delta \! y$
cannot be smaller than $\hbar$ due to the Heisenberg's uncertainty principle. 
We obtain, therefore, the following inequality:
\begin{equation}
p_{y}^\mathrm{ch}  d \gg \delta \! p_{y}  \delta \! y > \hbar .
\label{Heisenberg}
\end{equation}
The transverse motion of the projectile can be described by the laws of 
dynamics of a nonrelativistic particle with the mass $E/c^{2}$.
Here $E$ is the projectile energy and $c$ is the speed of light. 
The transverse kinetic energy of the channelling particle cannot exceed the depth of the
interplanar potential well $U_{\max}$:
\begin{equation}
\frac{(p_{y}^\mathrm{ch})^{2}}{2 E/c^{2}} < U_{\max}.
\label{kinenergy}
\end{equation}
 Combining (\ref{Heisenberg}) and
(\ref{kinenergy}) one sees that projectile can be considered classically if
its energy $E$ is sufficiently large:
\begin{equation}
E \gg \frac{(\hbar c)^{2}}{2 d^{2} U_{\max}} .
\label{totenergy_qq}
\end{equation}
Putting $U_{\max}=20$ eV and $d=1$ \AA\  one obtains $E \gg 0.1$ MeV.
However, 
the sign $\gg$ in this inequality has to be understood as 
'at least three orders of magnitude larger'. To show this, let us consider
a simple example of a parabolic potential well of depth $U_{\max}$ and width $d$:
\begin{equation}
U(y) = U_{\max} \left( \frac{y}{d/2} \right)^{2} .
\label{parabola}
\end{equation}
The oscillation frequency in such potential for a particle with the  
mass $E/c^{2}$ is $\omega = c \sqrt{U''(0)/E}$, where $U''$ is the second
derivative of the potential energy with respect to $y$. From (\ref{parabola}) one  
finds $U''= 8 U_{\max}/d$.
The number of quantum levels in the potential well is
\begin{equation}
n \approx \frac{U_{\max}}{\hbar \omega} = \frac{d}{2 \hbar c} \sqrt{\frac{U_{\max} E}{2}} .
\label{nlevels}
\end{equation}
Solving (\ref{nlevels}) for $E$ yields
\begin{equation}
E \approx 16 n^2  \frac{(\hbar c)^{2}}{2 d^{2} U_{\max}} .
\label{totenergy_approx}
\end{equation}
The right hand sides of (\ref{totenergy_qq}) and (\ref{totenergy_approx}) coincide
up to the factor of $16 n^{2}$.
The classical description is valid if the number of levels is sufficiently 
large. Taking $n=10$ one obtains $16 n^{2} =1600$. 

Hence, our model can be always applied to ultrarelativistic 
heavy projectiles. It is applicable in the case of light projectiles
if their energy is in the hundred MeV range or higher. The applicability 
conditions for electron projectile may be
somewhat stricter than for positrons, because the planar potential well is narrower 
in the case of negative particles.  
In the present paper we apply the model to electron projectiles at energy $E = 855$ MeV.
In this case the classical approximation is expected to be satisfactory.

The process of radiation  emission by the projectile is also treated classically.
The energy $\mathcal{E}$  
per unit interval of radiation frequency $\omega$
per unit of solid angle $\Omega$ 
emitted by the projectile is calculated according to formula (14.65) from \cite{Jackson_book}:
\begin{equation}
\frac{d^{3} \mathcal{E}}{d \omega d^2 \Omega} = 
\frac{q_\mathrm{p}^2}{4 \pi^2 c}
\left |
\int_{t_\mathrm{in}}^{t_\mathrm{out}} \! \! \! \mathrm{d} t
 \frac{\vec{\mathfrak{n}} \times [
(\vec{\mathfrak{n}}-\vec{\beta(t)}) \times \dot{\! \vec{\beta}}(t)
] }
{( 1- \vec{\mathfrak{n}} \cdot \vec{\beta}(t) )^2}
\exp \! \left \{  \mathrm{i} \, \frac{\omega}{c} \, [ ct - \vec{\mathfrak{n}} \cdot \vec{r}(t) ] \! \right \}
 \right   |^{2} \! \! .
\label{spectrum_int}
\end{equation}
Here $\vec{\mathfrak{n}}$ is a unit vector pointing from the crystal
to a distant observation point, $\vec{r}(t)$ is the projectile coordinate as a function of 
time, $\vec{\beta}(t)$ and $\dot{\! \vec{\beta}}(t)$ are respectively its velocity and acceleration
divided by the speed of light: $\vec{\beta}(t) = \dot{\! \vec{r}}(t)/c$, \ \ 
$\dot{\! \vec{\beta}}(t) = \ddot{\! \vec{r}}(t)/c$.
The integration over the time $t$ is taken from the moment $t_\mathrm{in}$ of entering to the 
moment $t_\mathrm{out}$ of existing the crystal by the projectile.

The classical approach is valid for relatively soft radiation: the photon energy has to be much smaller
than the energy of the projectile: $\hbar \omega \ll E$.

\section{Description of the Algorithm}
\label{Code}

The code performs 3D simulation of the motion of ultrarelativistic charged particles 
in a single crystal.

The crystal lattice is modelled as a collection of nodes. Each node is represented by 
the radius vector of its position and 
a few vectors, called `bonds', connecting the node to all its nearest neighbours.

In the case of diamond-type lattice, each node has four nearest neighbours. 
The bond length $b$ is calculated according to the formula 
\begin{equation}
b=\frac{\sqrt{3}}{2} v^{1/3},
\end{equation}
where $v$ of the volume per atom:
\begin{equation}
v = \frac{A}{\rho N_{\mathrm{A}}}.
\end{equation}
Here $A$ is the atomic weight, $\rho$ is the density of the crystal and $N_\mathrm{A}$
is the Avogadro constant.

The lattice is built as follows. A node is placed at the origin of the coordinate
system. The `bonds' of this node are oriented as described below. Then the lattice is `grown' by placing
new nodes at free ends of the `bonds' until the necessary space region is filled with the lattice.

To study the channelling along the crystallographic plane defined by the 
Miller indices $(klm)$ the lattice has to be oriented in such a way that 
its $(klm)$ plane  is parallel to the coordinate plane $(xz)$, where $z$ is the beam direction. 
This is accomplished by the proper orientation of the `bonds' of the initial node.
At the beginning, the `bonds' of the initial node are
\begin{eqnarray}
\vec{b}_{1} &=& \left( 1/\sqrt{3}, 1/\sqrt{3}, 1/\sqrt{3} \right)   \label{b1}\\
\vec{b}_{2} &=& \left( 1/\sqrt{3}, -1/\sqrt{3}, -1/\sqrt{3} \right) \label{b2}\\
\vec{b}_{3} &=& \left( -1/\sqrt{3}, 1/\sqrt{3}, -1/\sqrt{3} \right) \label{b3}\\
\vec{b}_{4} &=& \left( -1/\sqrt{3}, -1/\sqrt{3}, 1/\sqrt{3} \right) \label{b4}.
\end{eqnarray}
It corresponds to the orientation of the crystallographic directions $[100]$, $[010]$ and $[001]$
along the axes $x$, $y$ and $z$, respectively.
Then a rotation matrix $O_{(klm)}$ is built that transforms the vector (k,l,m) into the vector of the same length 
directed along the $y$-axis. All four `bonds' (\ref{b1})--(\ref{b4}) are multiplied by this matrix: 
$\vec{b}_{i} \rightarrow O_{(klm)} \vec{b}_{i}$, $i=1,2,3,4$.
Finally the `bonds' are also multiplied by the rotation matrix $O_{\phi}$ describing rotation around the $y$-axis 
by the angle $\phi$. The angle $\phi$ has to be carefully chosen to avoid collinearity of major crystal 
axes with the coordinate axis $z$.

In a similar way, a crystal axis can be oriented along the beam direction $z$ to simulate axial channelling.

There is also a possibility of a random orientation of the crystal avoiding incidental 
orientation of major crystal directions along the beam axis. In this case the crystal scatters
particles as an amorphous medium. This regime can be used to calculate the incoherent bremsstrahlung.

Once the lattice is constructed, the atoms are placed in its nodes.
Each atom is a collection of electric charges: a nucleus with the 
charge $+Ze$ and $Z$ electrons with charges $-e$, $Z=14$ in the case of silicon.
The positions of the electrons are random. Their distribution is spherically 
symmetric and 
the distance of electrons from the nucleus is calculated as
\begin{equation}
r_{i} = a_\mathrm{TF} \hat{r}_{i}, \hspace{2em} i=1,\dots,Z .
\end{equation}
Here, $a_\mathrm{TF}$ is the Thomas-Fermi radius of the atom:
\begin{equation}
a_\mathrm{TF} = \frac{0.8853}{Z^{1/3}}  a_\mathrm{B},
\end{equation}
with $a_\mathrm{B}$ being Bohr's radius, and $\hat{r}_{i}$ is found
by solving the following transcendental equation
\begin{equation}
\chi(\hat{r}_{i})-\hat{r}_{i} \chi'(\hat{r}_{i}) = \xi_{i}
\label{eqhatr}
\end{equation}
with $\xi_{i}$ being a random variable uniformly distributed within the
interval $0 < \xi_{i} \leq 1$.  
The function $\chi(\hat{r})$ in (\ref{eqhatr}) is the screening
function of the atomic potential. It can be shown that, 
if the positions of electrons are chosen as described above,
the total 
electrostatic potential of the nucleus 
and the electrons being averaged over the random positions of the electrons
has the form 
\begin{equation}
\langle U(r) \rangle = \frac{Z e}{r} \chi \left( \frac{r}{a_\mathrm{TF}} \right).
\end{equation}
The screening function $\chi(\hat{r})$ satisfies the conditions 
\begin{equation}
\xi(0)  =  1 , \hspace{3em}   \hspace{3em}
\lim_{\hat{r} \rightarrow +\infty} \chi(\hat{r})  =  0 ,
\end{equation}
so that the potential has the Coulomb form $(Z e)/r$  in the vicinity of the 
nucleus but is fully screened out by the electron cloud at large $r$.

We use the Moli\`ere screening function \cite{Moliere}, which has the form
\begin{equation}
\chi(\hat{r}) = \sum_{j=1}^{3} \alpha_{j} \exp \left( - \beta_{j} \hat{r} \right)
\end{equation}
with numerical parameters having the values
\begin{equation}
\begin{array}{lll}
\alpha_{1} = 0.35, &  \alpha_{2} = 0.55, &  \alpha_{3} = 0.1, \\
\beta_{1} = 0.3,   &   \beta_{2} = 1.2,  &  \beta_{3} = 6.0.
\end{array}
\end{equation}

To take into account quantum and thermal oscillations of the atoms in the crystal,
 the atomic nucleus is placed not exactly in the lattice node but is shifted 
from it by a random vector $\vec{\rho}$.
Each component of the random vector is normally distributed:
\begin{equation}
w({\rho}_{k}) = \frac{1}{\sqrt{2 \pi } a(T)} \exp \left[ - \frac{1}{2} \left( \frac{{\rho}_{k}}{a(T)} \right)^2  \right],
\hspace{2em} k=x, y, z
\label{ThermalVibrations}
\end{equation}
Here $a(T)$ is the average oscillation amplitude. We used the value $a(T) = 0.075$~\AA \ in our calculations, which
corresponds to the room temperature \cite{Crystallography}.

The trajectory of the projectile is modelled as follows:
The particle moves along straight line segments between the points where its 
coordinate $z$ coincides with the $z$ coordinate of one of a crystal constituent: 
an electron or a nucleus. At this point the transverse momentum of the projectile 
is changed according to equation (\ref{pperp}).
Then the projectile is moved further along a straight line segment corresponding to the 
transverse momentum $\vec{p}_{\perp} + \Delta \vec{p}_{\perp}$ until its $z$ coordinates coincides
with that of the next crystal constituent and a new modification of the particle momentum is performed.
 
\begin{figure}[htp]
\begin{center}
\includegraphics*[width=13.5cm]{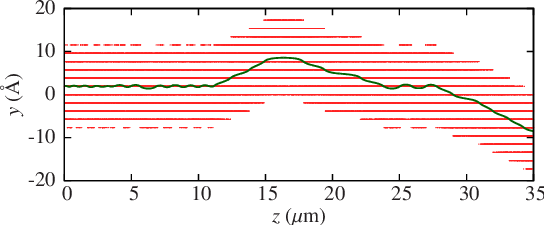}
\end{center}
\caption{An example of the crystal `pipe' surrounding the particle trajectory as it is modelled by the code.
A projection on plane $(zy)$ is shown.}
\label{crystal_cylinder}
\end{figure}

A crystal constituent is taken into account if it belongs to the lattice node located within 
a cylinder of the radius $40 a_\mathrm{TF}$ around the particle. Initially, the axis of the cylinder is the straight line along
the direction of the projectile momentum at the point of entering the crystal. The length of the cylinder 
is approximately 200 \AA. When the particle approaches the end of the cylinder, a new cylinder is built as an extension of the 
old one but along the direction of the new particle momentum. The procedure continues  
until the end of the crystal is reached.
As a result, the cylinders form a `pipe' filled by the crystal lattice and the particle channels inside it
as it is shown in 
Figure \ref{crystal_cylinder}. 
The computer time is saved substantially
due to the fact that only the part of the crystal lattice inside the `pipe' is 
modelled and the rest of it is ignored. The simulation takes
about 0.8 seconds per 1 $\mu$m of the particle trajectory on a 3 GHz CPU core. 

After the projectile trajectory is simulated, the radiation spectrum is calculated. 
The integral  in (\ref{spectrum_sum}) is approximated with a sum over the trajectory points:
\begin{equation}
\frac{\mathrm{d}^{3} \mathcal{E}}{\mathrm{d} (\hbar \omega) \mathrm{d}^2 \Omega} = 
\frac{\alpha}{4 \pi^2}
\left | \sum_{j} 
\left ( \! \frac{\delta \vec{\beta}_{j}}{D_{j-1}} + 
\frac{(\vec{\beta}_{j} - \vec{\mathfrak{n}}) (\delta \vec{\beta}_{j} \cdot \vec{\mathfrak{n}})}{D_{j} D_{j-1}} 
\! \right )
\exp \!
\left ( \! \mathrm{i} \, \frac{\phi_{j-1}+\phi_{j}}{2} \! \right )  \right |^{2} \! \! \!,
\label{spectrum_sum}
\end{equation}
where $\delta \vec{\beta}_{j}$ is an increment of $\vec{\beta}$ between two successive
trajectory points: $\delta \vec{\beta}_{j} = \vec{\beta}_{j} - \vec{\beta}_{j-1}$ 
and $\alpha$ is the fine structure constant.
The denominators $D_{j}$ are found from the formula $D_{j} = 1 - (\vec{\beta}_{j} \cdot \vec{\mathfrak{n}})$
and the phases $\phi_{j}$ are given by the expression 
\begin{equation}
\phi_{j} = \frac{\omega}{c} \, [ c t_{j} - \vec{\mathfrak{n}} \cdot \vec{r}_{j} ] .
\end{equation}
Dividing (\ref{spectrum_sum}) by $\hbar \omega$ yields the formula for the number of photons.


\section{Simulations}

\begin{figure}[htp]
\begin{center}
\begin{tabular}{cc}
\multicolumn{2}{c}{\Large (111)} \\
\includegraphics*[height=5.3cm]{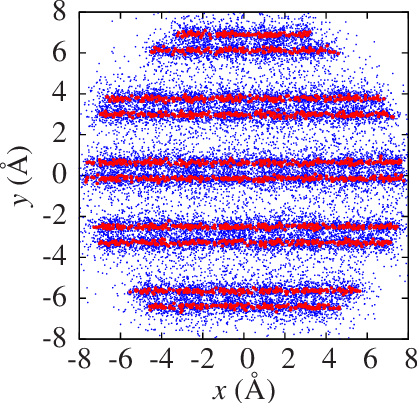} &
\includegraphics*[height=5cm]{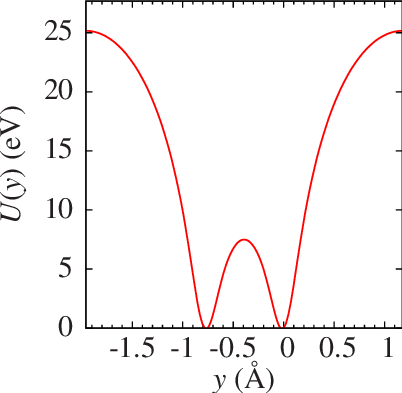} \\
\multicolumn{2}{c}{\Large (110)} \\
\includegraphics*[height=5.3cm]{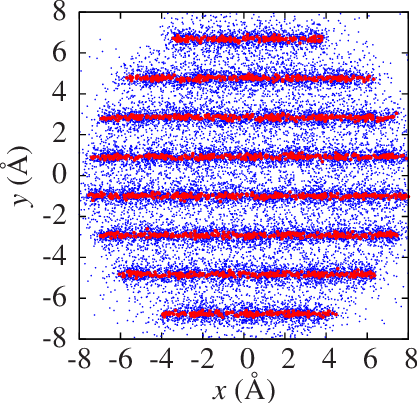} &
\includegraphics*[height=5cm]{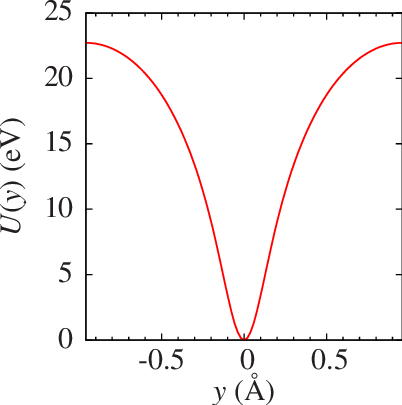} \\
\multicolumn{2}{c}{\Large (100)} \\
\includegraphics*[height=5.3cm]{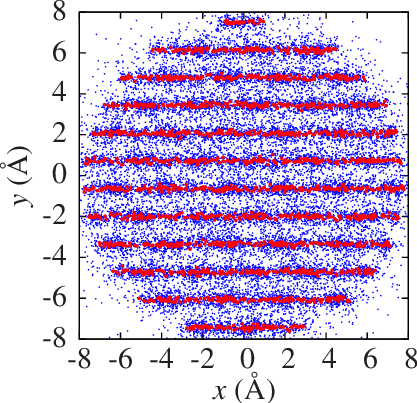} &
\includegraphics*[height=5cm]{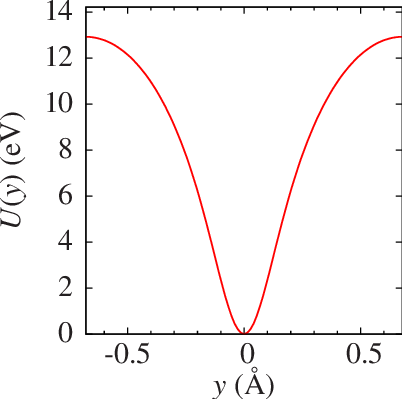} 
\end{tabular}
\end{center}
\caption{Left column: The projection of a 200 \AA \ long simulated  crystal 
cylinder on the (xy) plane. Smaller and larger circles stand respectively for electrons and nuclei.
Right column: the potential energy of the projectile electron in the field of crystal planes in
continuous approximation.
The plots in right column serve an illustrative purpose only, the continuous potentials
are not used in the calculation algorithm.}
\label{ProfilesAndPotentials}
\end{figure}

The calculations were performed for E=855 MeV electrons in a single crystal of silicon for three crystal 
orientations corresponding to channelling along (100), (110) and (111) planes. The simulated positions
of the crystal constituents and the potential energy of the projectile electron in the electrostatic field
of crystal planes
are shown in Figure \ref{ProfilesAndPotentials}.

Initially, the projectiles had zero transverse momentum. This corresponds to the 
ideal case of a zero-emittance beam entering the crystal strictly parallel to the coordinate axis $z$. 
The transverse position of the projectile
at the entrance of the crystal was chosen randomly, homogeneously distributed
along the channel width. Then the trajectory of the particle was simulated
as it is described in the previous section. The simulation of the trajectory was
terminated if the particle went through the crystal: $z > L_\mathrm{cr}$, or if the 
deviation of the projectile from its initial direction became too large: 
$| \vec{p}_{\perp} | / p_{z} > 100/\gamma$ (here $p_{\perp}$ and $p$ are respectively the 
transverse and the longitudinal momenta and $\gamma$ is the Lorentz factor of the projectile.)

Each simulated trajectory was analysed to determine the segments  corresponding 
to the channelling and dechannelled regime. The particle was considered to be in 
the channelling regime from the point of entering the crystal to the point were 
its crossed one of the two channel boundaries. Then the particle was
considered to be dechannelled. If a dechannelled particle changed the direction 
of the $y$ component of its velocity two or more times without crossing channel
boundaries (i.e. if it made at least one complete channelling 
oscillation  within a channel) it was considered to be rechannelled and remaining in the channelling 
regime until it crossed one of the channel boundaries again.

The number of the simulated trajectories for the analysis of dechannelling was 40000, 30000 and 
about 28000 for planar channels (111), (110) and (100), respectively.
The dimension of the crystal in the beam direction $L_\mathrm{cr}$ was equal to maximum length of the crystals 
used in recent channelling experiments \cite{Backe2008}: $L_\mathrm{cr}=270.4$ $\mu$m.  

Channelling radiation was calculated for the plane (110) and for 'amorphous' orientation (i.e. for a crystal 
oriented randomly avoiding major crystal directions) for seven different values of $L_\mathrm{cr}$ ranging from
$L_\mathrm{cr}=7.9$ $\mu$m to $L_\mathrm{cr}=270.4$ $\mu$m, 50000 trajectories were simulated in each case.

\section{Definition of the Dechannelling Length}
\label{definitionLd}

To make a quantitative assessment of the particle dechannelling process, one needs a definition 
of the dechannelling length that would be suitable for the Monte Carlo approach.

Let $z_\mathrm{d1}$ be the point of the first dechannelling of the projectile.
We define the quantity $N_\mathrm{ch0}(z)$ as the number of projectile particles
for which $z_\mathrm{d1} > z$, i.e. this is the number of particles that passed the distance 
from the crystal entrance to the point $z$ in the channelling regime and dechannel 
at some further point. The length $L(z)$ is the average distance from the point $z$ to the 
first dechannelling point:
\begin{equation}
L(z) = \frac{\sum_{k=1}^{N_\mathrm{ch0}(z)} (z_\mathrm{d1}^{(k)} - z) }{N_\mathrm{ch0}(z)} .
\label{Lz}
\end{equation}
The sum in the numerator is taken over those projectiles for which $z_\mathrm{d1} > z$.

Generally speaking, $L(z)$ depends not only on $z$, but also on the angular distribution 
of the particles at the crystal entrance. Nonetheless, as it will be shown below, 
the kinetic theory of channelling suggests that,
at sufficiently large $z$, $L(z)$ reaches an asymptotic 
value  that depends neither on $z$ nor on the initial angular distribution. 

From the solution of the diffusion equation (see, for instance, formula (1.38) in 
\cite{BiryukovChesnokovKotovBook}),  one can obtain the following expression for
$N_\mathrm{ch0}(z)$
\begin{equation}
N_\mathrm{ch0}(z) =N_{0} \sum_{j=1}^{\infty} A_{j} \exp \left(- z / L_{j} \right).
\label{Nch0_diff}
\end{equation}
Here only coefficients $A_{j}$ depend on the initial angular distribution of the particles, 
while the lengths $L_{j}$ depend exclusively on the properties of the crystal channel and the
energy, charge and mass of the projectile. 

The $1/\mathrm{e}$ dechannelling length $L_\mathrm{d}$ is defined as the largest 
of the parameters $L_{j}$ in (\ref{Nch0_diff}). The corresponding term dominates the 
asymptotic behaviour at $z \gtrsim L_\mathrm{d}$:
\begin{equation}
N_\mathrm{ch0}(z) \asymp N_{0} A_{d} \exp \left(- z / L_\mathrm{d} \right) .
\label{Nch0_diff_asymp}
\end{equation}

The expression (\ref{Lz}) for $L(z)$ has the following counterpart in the kinetic theory
\begin{equation}
L(z) = - \frac{1}{N_\mathrm{ch0}(z)} \int_{z}^{\infty}  \mathrm{d}  z_\mathrm{d1} (z_\mathrm{d1} - z) 
\frac{\mathrm{d} N_\mathrm{ch0}(z_\mathrm{d1})}{\mathrm{d} z_\mathrm{d1}} 
\label{Lz_diff}
\end{equation}
Substituting (\ref{Nch0_diff_asymp}) into (\ref{Lz_diff}) demonstrates that, indeed, the coefficient $A_{d}$
cancels out and L(z) becomes equal to $L_\mathrm{d}$ in the asymptotic region. 

Although the diffusion equation was solved in \cite{BiryukovChesnokovKotovBook} for positively charged projectile
in harmonic potential approximation, the exponential asymptotic behaviour of $N_\mathrm{ch0}(z)$, and, consequently,
a constant asymptotic value of $L(z)$  is a more general result. As it will be shown in the 
next section, our simulations demonstrate that it is also valid for electrons.

Hence, in our Monte Carlo procedure the dechannelling length $L_\mathrm{d}$ is defined as the asymptotic value of 
$L(z)$ in the region where it ceases to depend on $z$.

\section{Analysis of the Results}

\begin{figure}[htp]
\begin{center}
\includegraphics*[width=12cm]{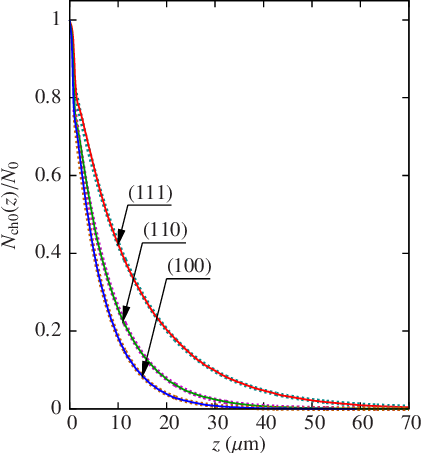}
\end{center}
\caption{Fraction $N_\mathrm{ch0}(z)/N_{0}$
of the particles that stay in the same channel from their entrance into the crystal for different 
planar channels as function of the penetration depth $z$. 
The thick dashed lines show the corresponding exponential asymptotes $\propto \exp(-z/L_\mathrm{d})$, the values
of $L_\mathrm{d}$ are listed in Table \ref{Ld_table}.}
\label{ch0}
\end{figure}

The ratio $N_\mathrm{ch0}(z)/N_{0}$ 
as function of  $z$
is shown in 
Figure \ref{ch0}  
 for three different crystal channels.
This fraction decreases rather fast and, as it was expected, has an exponential  asymptotic behaviour.

\begin{figure}[htp]
\begin{center}
\includegraphics*[width=12cm]{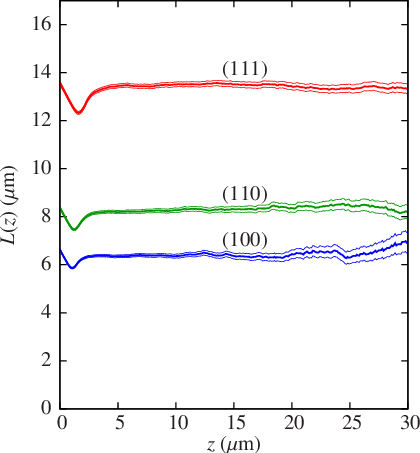}
\end{center}
\caption{The quantity $L(z)$ (\ref{Lz}) that becomes equal to the dechannelling length
at large penetration depth $z$. The thin lines show the statistical errors.}
\label{Ld}
\end{figure}

The quantity $L(z)$ (\ref{Lz}) for the same channels is plotted in Figure \ref{Ld}.
Indeed, $L(z)$ becomes constant (within the statistical errors) at large $z$ corresponding 
to exponential behaviour 
of the curves of Figure \ref{ch0}. The asymptotic values, $L_\mathrm{d}$, are listed
in Table \ref{Ld_table}.

\begin{table}
\caption{Monte Carlo results for the $1/\mathrm{e}$ dechannelling length $L_\mathrm{d}$ for three different crystal channels.
The results for initial beam and rechannelled particles coincide within the statistical errors.}
\label{Ld_table}
\begin{center}
\begin{tabular}{|c|c|c|}
\hline
Crystal & \multicolumn{2}{|c|}{Dechannelling Length}\\
Plane & \multicolumn{2}{|c|}{($\mu$m)} \\ \cline{2-3}
& \hspace*{2em}  Initial beam \hspace*{2em} & Rechannelled particles  \\
\hline
(111) & $ 13.57 \pm 0.12 $ & $13.69 \pm 0.07 $\\
(110) & $ 8.26 \pm 0.08  $ & $ 8.38 \pm 0.05 $\\
(100) & $ 6.38 \pm 0.07  $ & $ 6.40 \pm 0.05 $\\
\hline
\end{tabular}
\end{center}
\end{table}

\begin{figure}[htp]
\begin{center}
\includegraphics*[width=12cm]{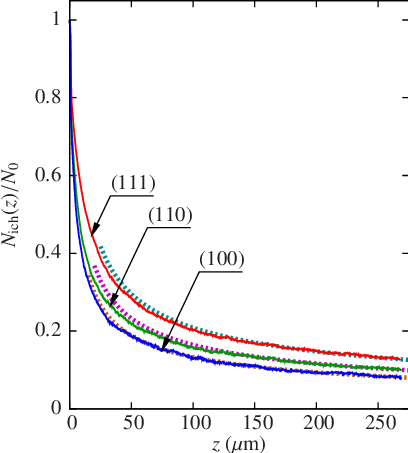}
\end{center}
\caption{The fraction of channelling particles as function of penetration depth $z$ for different 
crystal channels (solid lines). The thick dashed lines show the corresponding asymptotes $\propto z^{-1/2}$. }
\label{ich}
\end{figure}

Only the particles that remained in the channelling regime 
from their entrance to the crystal were considered in Figures \ref{ch0} and  \ref{Ld}.
The fraction of these particles
decreases fast. In contrast, the fraction $N_\mathrm{ich}(z)/N_{0}$
of the particles that are in the 
channelling regime at the point $z$ regardless of their previous channelling status
decreases  rather slowly (see Figure \ref{ich}). The reason for it is the rechannelling process.
Random collisions with the crystal constituents can occasionally reduce the transverse 
energy 
\begin{equation}
E_y = \frac{p_y^2}{2 E/c^2} + U(y).
\label{Ey}
\end{equation}
Therefore, a dechannelled particle can return to the channelling regime. 

Rechannelling occurs more often for electrons than for positively charged particles. 
This
is due to the fact that the random scattering is more intense
in the vicinity of the crystal plane. For positively charged particles this means that 
the scattering is most probable at the top of the potential barrier,
i.e. near the maximum of the potential energy $U(y)$. 
Even if a collision 
happens to reduce the component $p_y$  of the projectile momentum 
to zero, the transverse energy $E_y$
still remains in the vicinity of the top of the potential barrier. The range of $p_y$
at which the particle returns to the channelling regime is zero at the maximum of the 
potential and is small in the vicinity of it.
 Therefore,
probability of the rechannelling is small.

In contrast, the potential minimum for negatively charged particles is located near the 
crystal planes. This means that the random collisions are most probable near the 
minimum of the potential energy,
where there is a wider range of $p_y$ at which $E_y$ drops below the
potential barrier. Hence, the probability of rechannelling is higher for negative
than for positive particles.

\begin{figure}[htp]
\begin{center}
\includegraphics*[width=12cm]{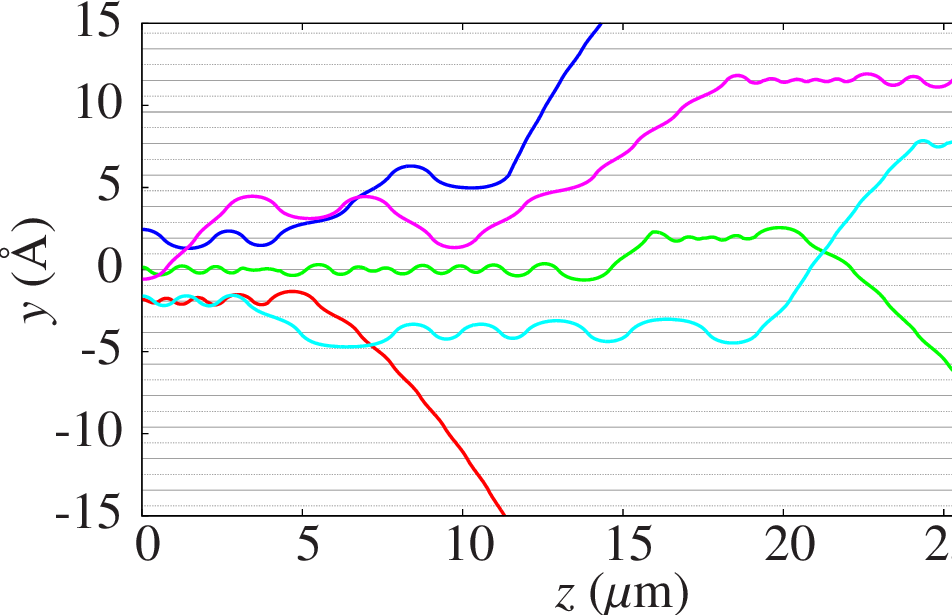}
\end{center}
\caption{Examples of simulated trajectories. Only initial segments corresponding to
$z < 30$ $\mu$m are plotted. The crystallographic planes are shown by solid horizontal lines,
dashed lines show the boundaries between the channels (the maxima of the interplanar potential).
Four of five trajectories demonstrate rechannelling. One of the
particles rechannels twice.}
\label{traj}
\end{figure}

Typical trajectories of electrons are shown in Figure \ref{traj}, where
rechannelling is clearly seen. According to our results, a particle rechannels 
in average 4.8 times on the length of the crystal $L_\mathrm{cr}=270.4$ $\mu$m
into the channel (100). For the channels (110) and (111) the corresponding numbers are 
respectively 4.4 and 3.5.

The asymptotic behaviour of the curves in Figure \ref{ich} can be explained in the following way.
At sufficiently large $z$, the distribution of the dechannelled particles with respect to the transverse momentum 
$p_y$ is similar to that in an amorphous medium and can be approximated by the Gaussian function:
\begin{equation}
w(p_y) = \frac{1}{\sqrt{2 \pi } \sigma(z)} \exp \left[ - \frac{1}{2} \left( \frac{p_y}{\sigma(z)} \right)^2  \right]
\label{Gauss}
\end{equation}
with the variance proportional to $z$:  $\sigma^{2}(z) \propto z$. The rechannelling is dominated 
by the phase space density the vicinity of the point $p_y=0$. It decreases as 
$1/\sigma(z) \propto 1/\sqrt{z}$ and
governs the asymptotic behaviour of the fraction of the channelling particles shown in Fig. \ref{ich}.

For the analysis of the rechannelled particles, one can consider the quantity
$N_{\mathrm{ch}n}(\tilde{z})$, which is the number of particle that rechannelled
at least $n$ times and travelled at least the distance $\tilde{z}$ from the 
$n$-th rechannelling point $z_{\mathrm{r}n}$ in the channelling regime, i.e. 
this is the number of particles for which
\begin{equation}
z_{\mathrm{d}(n+1)}-z_{\mathrm{r}n} > \tilde{z},
\label{defNchn}
\end{equation}
where $z_{\mathrm{d}(n+1)}$ is the point of $(n+1)$-th dechannelling.
Note that the longer is the crystal along the beam direction the larger 
is the probability of 
rechannelling, therefore $N_{\mathrm{ch}n}(\tilde{z})$ for $n>0$ depend 
on $L_\mathrm{cr}$, in contrast to $N_{\mathrm{ch0}}$.

It is reasonable to expect that the particle `forgets' the value of 
its initial transverse momentum by the point of its first rechannelling or even earlier. 
Therefore, all the rechannellings 
($1$st, $2$nd, and so on) are expected to be statistically identical and it 
makes sense to analyse them together. We introduce the quantity
\begin{equation}
N_{\mathrm{rch}}(\tilde{z}) = N_{\mathrm{ch1}}(\tilde{z}) + N_{\mathrm{ch2}}(\tilde{z}) 
+ N_{\mathrm{ch3}}(\tilde{z}) + \dots
\label{defNrch}
\end{equation}
Note that $N_{\mathrm{rch}}(\tilde{z})$ may be larger than the total number of particles 
$N_{0}$, because each particle may rechannel several times.

\begin{figure}[htp]
\begin{center}
\includegraphics*[width=12cm]{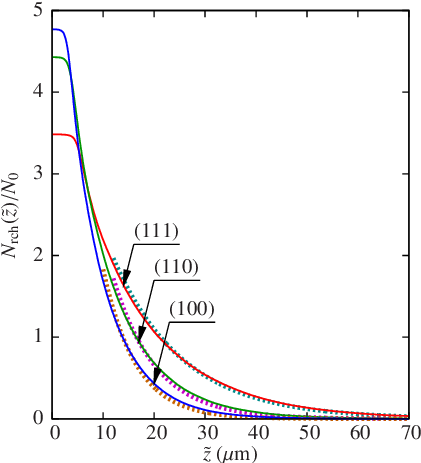}
\end{center}
\caption{The ratio  $N_{\mathrm{rch}}(\tilde{z})/N_{0}$ (see text)  for different 
planar channels as function of the distance $\tilde{z}$ from the rechannelling point.
The thick dashed lines show the corresponding exponential asymptotes $\propto \exp(-z/L_\mathrm{d})$, the values
of $L_\mathrm{d}$ are listed in Table \ref{Ld_table}.}
\label{rch}
\end{figure}

The ratio $N_{\mathrm{rch}}(\tilde{z})/N_{0}$ for $L_\mathrm{cr}=270.4$ $\mu$m for 
three crystal channels is plotted in Figure \ref{rch}. Similarly to $N_{\mathrm{ch0}}(z)$,
this quantity decreases fast with $\tilde{z}$ and has an exponential asymptote.

It is possible to introduce the quantity $L(\tilde{z})$, which is a counterpart of (\ref{Lz})
for rechannelled particles:
\begin{equation}
L(\tilde{z}) = \frac{\sum_{k} \sum_{n} (z_{\mathrm{d}(n+1)}^{(k)}-z_{\mathrm{r}n}^{(k)} - \tilde{z}) }{N_{\mathrm{rch}}(\tilde{z})}.
\label{Lzr}
\end{equation}
The first sum in the numerator runs over all particles that rechannelled at least once. The second sum runs over all rechannellings of each 
particle.

\begin{figure}[htp]
\begin{center}
\includegraphics*[width=12cm]{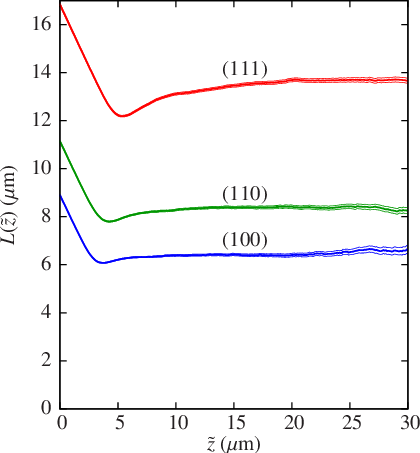}
\end{center}
\caption{The quantity $L(\tilde{z})$  (\ref{Lzr})
for different 
planar channels as function of the distance $\tilde{z}$  from the rechannelling point.
The thin lines show the statistical errors.
Note that the asymptotic values of $L(\tilde{z})$ are the same as for $L(z)$ (cf. 
Figure \ref{Ld}).
}
\label{Ldr}
\end{figure}

The quantity $L(\tilde{z})$ is plotted in Figure \ref{Ldr}.
The behaviour of $L(\tilde{z})$ 
and $L(z)$  (see Figure \ref{Ld}) at small values of $\tilde{z}$ and $z$ is
different, because the transverse momentum distribution of the rechannelling particles 
is quite different from that of the ideally parallel initial beam. 
But asymptotic behaviour of both quantities is essentially the same: they both 
reach a constant value, which is, by definition, the dechannelling 
length $L_\mathrm{d}$.  In Table \ref{Ld_table}, the values of $L_\mathrm{d}$ 
obtained from the analysis of 
rechannelled particles is compared to those obtained for the initial beam. They 
coincide within the statistical errors. This confirms that the dechannelling 
length calculated according to our definition does not depend on the initial 
transverse momentum distribution of the projectiles.

\section{Comparison to Experiment}

The dechannelling length cannot be measured directly in an experiment because it is not 
possible to separate the particles that were in the channelling regime 
from the entrance point and the rechannelled particles.
Only signals related
to the total number 
of channelling  particles can be measured. 
Extracting the dechannelling length
from these data involves a model-dependent procedure. Therefore, comparing 
the values of $L_\mathrm{d}$ obtained by Monte Carlo simulations
to estimations found in the experimental publications would be a comparison of two theoretical 
models rather than an experimental verification of the code.

A correct way to check a physical model and the corresponding computer code is to use it 
for calculation of those quantities that can be directly measured in an experiment. Then
these results should be compared to the experimental data.

In the experiment at Mainz Microtron \cite{Backe2008}
the intensity of the channelling radiation was measured for crystal samples of 
different dimensions $L_\mathrm{cr}$ along the beam axis. To make a comparison with these data, we
modelled the Mainz experiment with our code.

We calculated the average number of photons in the energy interval 
$0.4 $ MeV $< \hbar \omega < 9.0$ MeV emitted by a 855 MeV electron moving
through a Silicon
crystal with plane (110) parallel to the beam direction.
Then we subtracted the background, i.e. the same quantity 
but calculated in the case of a randomly oriented Silicon crystal. 
The photons were taken into account if the angle $\theta$ between their wave vector 
and the beam direction, does not exceed $1.31$ mrad, which corresponds to the aperture 
of the gamma spectrometer in the experimental setup of the Mainz experiment. 
The calculation were done for different values of $L_\mathrm{cr}$.

\begin{figure}[htp]
\begin{center}
\includegraphics*[width=12cm]{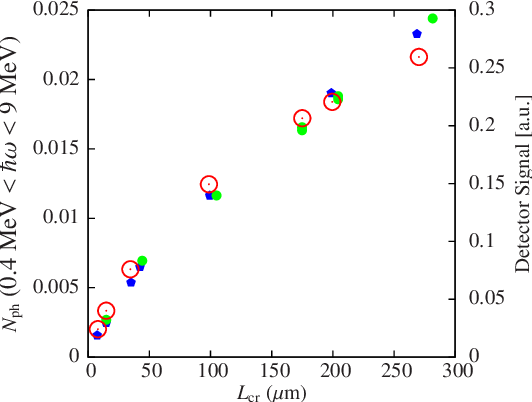}
\end{center}
\caption{Channelling radiation as function of crystal size along the beam 
direction. The filled symbols are experimental data \cite{Backe2008} and
the open cycles are results of our calculations. The left vertical axis shows 
the number of emitted photons per projectile within the energy interval
$0.4 $ MeV $< \hbar \omega < 9.0$ MeV. The right vertical axis is calibrated
in arbitrary units used in the experimental paper.}
\label{chan_rad}
\end{figure}

The intensity of the channelling radiation is presented in \cite{Backe2008} in arbitrary 
units. We equated $12$ arbitrary unit to $1$ photon per projectile to adjust the overall 
scale. The results are shown in Figure \ref{chan_rad}.
As is seen, our results demonstrate reasonable agreement with the experiment, which proves the 
reliability of the code.\footnote{There is, however, a discrepancy in the dechanneling length:
the value for the (110) channel estimated by the Mainz group $L_\mathrm{d}=18\ \mu$m  \cite{Backe2008}
is by the factor of more than two larger than our result $L_\mathrm{d}=8.26\ \mu$m.  It has to be pointed 
out that a model-dependent procedure was used in \cite{Backe2008} to estimate the $L_\mathrm{d}$.
Therefore, the discrepancy in the dechanneling length does not mean that there is disagreement of our results
with the experiment. The reasons for the discrepancy are still to be clarified. Preliminary, we attribute it 
to different definitions of the dechanneling length.}

\section{Conclusion and Discussion}

We presented first result obtained with a new Monte-Carlo code for modelling of
channelling of ultrarelativistic charged particles in a crystal. The calculation were
done for 855 MeV electrons channelling in a single crystal of Silicon along 
(100), (110) and (111) crystallographic planes. 

According to our simulation, if rechannelling is disregarded, the number of channelling 
electrons decreases fast with the penetration depth $z$ and quickly approaches 
an exponential asymptote.
Similar behaviour was previously seen in the kinetic theory of channelling in the 
case of positively charged projectile. 

We formulated a definition of the $1/\mathrm{e}$ dechannelling length $L_\mathrm{d}$ that 
is suitable
for application within the Monte Carlo approach. Our definition is consistent with the 
one previously used in the kinetic theory of channelling. Applying this definition to 
the initial beam and to rechannelled particles gives essentially the same result. This 
demonstrates that $L_\mathrm{d}$ is a universal quantity and does not depend on the 
initial transverse momentum distribution of projectiles.

We calculated the dechannelling length for the studied planar channels. It appeared
to be in the $10$ $\mu$m range.  

Our simulations show that the rechannelling of electrons is a notable phenomenon.
It  dominates the number of channelling particles already at the penetration depth 
of a few tens of microns. Due
to rechannelling, the total number of channelling particles, included the rechannelled fraction, 
decreases slowly following the $\propto 1/\sqrt{z}$ asymptote.

To verify our code, we calculated the intensity of the channelling radiation and 
compared it with the experimental data obtained at Mainz Microtron \cite{Backe2008}.
 A good agreement 
was observed. This confirms that our code is a reliable tool for modeling the
electron channelling if the projectile energy is around $1$ GeV.

There is no obstacle for successful application of the code also to positrons
of the same energy range or to electrons and positrons of higher energy up to 
several tens of GeV.
Extending the applicability domain of the code to much lower or to much higher 
projectile energies requires further improvements of the underlying 
physical model and the computation algorithm.

At low projectile energies (around $100$ MeV or lower)  the splitting of the transverse
energy levels of a channelling electron or positron
can be as large as a few electron-volts, which is comparable 
with the depth of the potential well. Because only discrete transverse energy levels are allowed, 
scattering from the crystal 
constituents will lead to an increase of the transverse energy 
only when the energy transfer in the collision is equal
to the splitting.
Hence, only collisions with 
large  scattering angles 
can contribute to the increase of transverse energy.
Due to smaller probability of such collisions, the dechannelling  length 
may become noticeably larger than it is predicted by classical calculations.
For these reasons, a study of low energy electron and positron channelling
would require taking into account quantum properties of the 
projectile.

At high electron or positron energies (hundreds of GeV or higher), the radiation energy losses 
become essential and cannot be ignored. Therefore, our model has to be further developed to take 
these into account.

In the case of heavy projectiles, the radiation energy losses are much less important. Therefore, in
principle, our model can be applied to heavy projectiles of TeV energy range or even 
higher. However, calculations of heavy projectile channelling may involve very long crystals because
of a large dechannelling length. This could require a prohibitive amount of computer time.
Therefore, further refinement and optimisation of the algorithm may be necessary in this case.

\ack
We are grateful to Hartmuth Backe, Werner Lauth and Dirk Krambrich for fruitful discussions.
Our work was supported by the Deutsche Forschungsgemeinschaft (DFG).

\end{document}